\documentclass[journal,comsoc]{IEEEtran}
\usepackage[noadjust]{cite}
\usepackage{float}
\usepackage{graphicx}
\usepackage[normalem]{ulem}

\usepackage{balance}
\usepackage{tabularx,booktabs}
\newcolumntype{Y}{>{\centering\arraybackslash}X}
\usepackage{multirow}
\usepackage[usenames,dvipsnames]{xcolor}
\usepackage{svg}

\usepackage[utf8]{inputenc}
\usepackage{graphicx}
\usepackage{enumitem}\setlist[description]{font=\textendash\enskip\scshape\bfseries}

\usepackage[most]{tcolorbox}
\pdfoutput=1
\usepackage{caption}
\usepackage{hyperref}
\usepackage{subcaption}
\title{Elastic Federated Learning over Open Radio Access Network (O-RAN) for Concurrent Execution of Multiple Distributed Learning Tasks}
\author{Payam Abdisarabshali, Nicholas Accurso,~\IEEEmembership{Student Member,~IEEE}, Filippo Malandra,~\IEEEmembership{Member,~IEEE},\\ Weifeng Su,~\IEEEmembership{Fellow,~IEEE}, and Seyyedali Hosseinalipour,~\IEEEmembership{Senior Member,~IEEE}}
\date{November 2022}

\begin{document}
\maketitle

\begin{abstract}
Federated learning (FL) is a popular distributed machine learning (ML) technique in Internet of Things (IoT) networks, where resource-constrained devices collaboratively train ML models while preserving data privacy. However, implementation of FL over 5G-and-beyond  wireless networks faces key challenges caused by (i) dynamics of the wireless network conditions and (ii) the coexistence of multiple FL-services in the system. In this paper, we unveil two key phenomena that arise from these challenges: \textit{over/under-provisioning of resources} and \textit{perspective-driven load balancing}, both of which significantly impact FL performance in IoT environments. We take the first steps towards addressing these phenomena by proposing a novel distributed ML architecture called \textit{elastic FL} ({\tt EFL}). {\tt EFL} unleashes the full potential of Open RAN (O-RAN) systems and introduces an elastic resource provisioning methodology to execute FL-services. It further constitutes a multi-time-scale FL management system that introduces three dedicated network control functionalities tailored for FL-services, including \textit{(i) non-real-time (non-RT) system descriptor}, which trains ML-based applications to predict both system and FL-related dynamics and parameters;  \textit{(ii) near-RT FL controller}, which handles O-RAN slicing and mobility management for the seamless execution of FL-services; \textit{(iii) FL MAC scheduler}, which conducts real-time resource allocation to the end clients of various FL-services. 
We finally prototype {\tt EFL} to demonstrate its potential in improving the performance of FL-services. { Our implementations are publicly available at the following GitHub repository: \href{https://github.com/payamsiabd/Elastic_FL_Over_O_RAN.git}{https://github.com/payamsiabd/Elastic\_FL\_Over\_O\_RAN.git}
}
\end{abstract}
\vspace{-4mm}
\section{Introduction}\label{sec:intro}
\noindent Federated learning (FL) has attracted tremendous attention for executing data-intensive Internet-of-Things (IoT) applications (e.g., smart manufacturing), where data is distributedly collected at edge devices~\cite{9060868}. This distributed machine learning (ML) approach is an alternative to centralized ML methods that require transferring distributed data to cloud servers, which often cause a high communication overhead and concerns regarding the exposure of raw data to untrusted parties. FL performs ML training through the repetition of multiple global training rounds, with each round consisting of two key steps: (i) using their local data, clients/devices perform \textit{local model} training, e.g., via several local iterations of stochastic gradient descent (SGD), and transfer their local models to a server, and (ii) the server aggregates all the received models (e.g., via averaging) to a \textit{global model} and then broadcasts it back to the clients to initiate the next global training round.

\vspace{-3mm}
\subsection{Motivation}
Despite its tremendous potential, FL is anticipated to be deployed on resource-constrained IoT wireless devices such as IoT sensors~\cite{9060868}, which is challenging due to the increasing size of modern ML models that demand substantial computational power for local model training and incur high communication overheads during the transfer from devices to the central server. 
To this end, research -- see \cite{9060868} and the references therein -- has focused on  (i) tuning the number of local SGD iterations and frequency/period of client-to-server communications; (ii) model compression, gradient sparsification, or selective parameter sharing; (iii) efficient client selection/recruitment. Despite the significance of these methods, there are two key underlying assumptions made in all of them, limiting their applicability in real-world scenarios:

\textbf{(A1) \textit{Snapshot-Based Analysis}:} These methods often assume that network and client conditions, such as client location and client-to-server channels, remain fixed during each global training round of FL, and thus they conduct \textit{snapshot-based analysis} of the network. Afterwards, in each snapshot -- often presumed to be the condition of the network at the beginning of each global training round -- they make control decisions (e.g., spectrum allocation and client recruitment). However, in real-world scenarios, such analysis may quickly become sub-optimal, as the network condition may vary over time due to factors like client mobility and evolving client datasets. 
    
\textbf{(A2) \textit{Isolated Execution of a Single FL-Service}:} The existing methods primarily focus on executing a single FL-service. However, in large-scale IoT networks, multiple FL-services may simultaneously compete for wireless resources and client recruitment (e.g., Google may run an FL-service for next-word prediction, while Apple does so for face recognition). Despite its importance, only a few studies have explored this topic~\cite{9799768,9445589}, all of which consider snapshot-based analysis suffering from the above limitations.

The simultaneous relaxation of the above assumptions (A1)\&(A2) unveils a new paradigm in which \textit{multi-service FL execution} intersects with \textit{dynamic wireless resource management}, introducing the following two non-trivial challenges:

\textbf{\textit{(1)  Over-/Under-Provisioning of Resources}:} 
The relaxation of \textbf{(A1)} introduces critical design considerations focused on the over-/under-provisioning of resources (e.g., wireless spectrum), stemming from two phenomena: (i) \textit{Signal Strength Fluctuation:} This issue is caused by client mobility and time-varying channels. For example, a client experiencing poor channel conditions at the start of a global round might initially be allocated high bandwidth to offset its weaker connection. However, as the client moves to a position with a better channel, the previously allocated high bandwidth may end up being underutilized; (ii) \textit{Incomplete Information:} This reflects a situation where FL-services lack complete information about each other's objectives, strategies, and the congestion of clients at different areas. This incompleteness often stems from network access restrictions, where FL-services are not granted access to view network information, such as bandwidth availability and client conditions. In this situation, due to selfishness of FL-service owners, they may make greedy decisions and occupying shared resources (e.g., spectrum), which can lead to over-provisioning of resources for some FL-services while inadvertently cause under-provisioning for others. 

\textbf{\textit{(2) Perspective-Driven Load Balancing}:}
Load balancing refers to distributing the network traffic, preventing  congestion across the network. 
The coexistence of multiple FL-services (relaxation of \textbf{(A2)})  calls for a unique load balancing strategy that considers a nuanced phenomenon which we call \textit{perspective-driven client distribution}. { This phenomenon captures the fact that the distribution of FL clients across different network regions varies depending on the perspective from which clients are evaluated/perceived. For example, 
regions with advanced connectivity infrastructure, such as urban areas, may have a higher density of clients suitable for latency-sensitive FL-services. Also, certain regions may have a higher density of low-cost clients, making them ideal for cost-sensitive FL-services, while other regions may have clients with greater computational capabilities, better suited for computation-intensive FL-services.}
Given the perspective-driven client distribution, in FL, load balancing must account for the overlap of the FL-services' objectives. 

\vspace{-3mm}
\subsection{Migration from Per-User to Service-Level Objectives}
Simultaneous addressing of the above two challenges calls for client mobility management (i.e., handover design for load balancing) and dynamic resource allocation solutions, which are the key \textit{network control functionalities} of radio access networks (RANs)~\cite{9076124, 8737604}. Additionally, orchestrating the coexistence of multiple FL-services over a dynamic IoT network, while addressing the above two challenges, demands \textit{a system management approach operating at different levels of agility}. For example, while client recruitment for different FL-services can be performed at the start of each global round, mobility management requires a finer-granular time scale (real-time) to accommodate the fast-paced changes in network topology due to client repositioning.
However, the existing IoT network control functionalities are often developed for native wireless services (e.g., high data-rate Internet provisioning) and primarily focus on \textit{per-user QoS requirements}, including energy consumption and communication latency. 
Instead, in FL, addressing the aforementioned two challenges should consider 
\textit{collective service-level objectives/characteristics, such as the prediction accuracy of the FL models and client recruitment costs, alongside  QoS metrics like overall energy consumption and model training latency across FL-services.} Additionally, effective load balancing in FL must ensure an optimal allocation of \textit{clients to FL-services and wireless resources to the clients}, while considering the innate differences among FL-services. For example, consider an FL-service for monitoring industrial equipments utilized during business hours coexisting with an FL-service focused on air quality monitoring. Equipment monitoring might require rapid ML model training to predict failures, while environmental sensing may tolerate longer ML training delays. In this case, the system should prioritize allocating network resources to industrial equipment monitoring during business hours and then shift resources towards environmental sensing.

\begin{figure*}
    \centering 
    \includegraphics[width=0.55\linewidth,trim=0 0 0 0,clip]{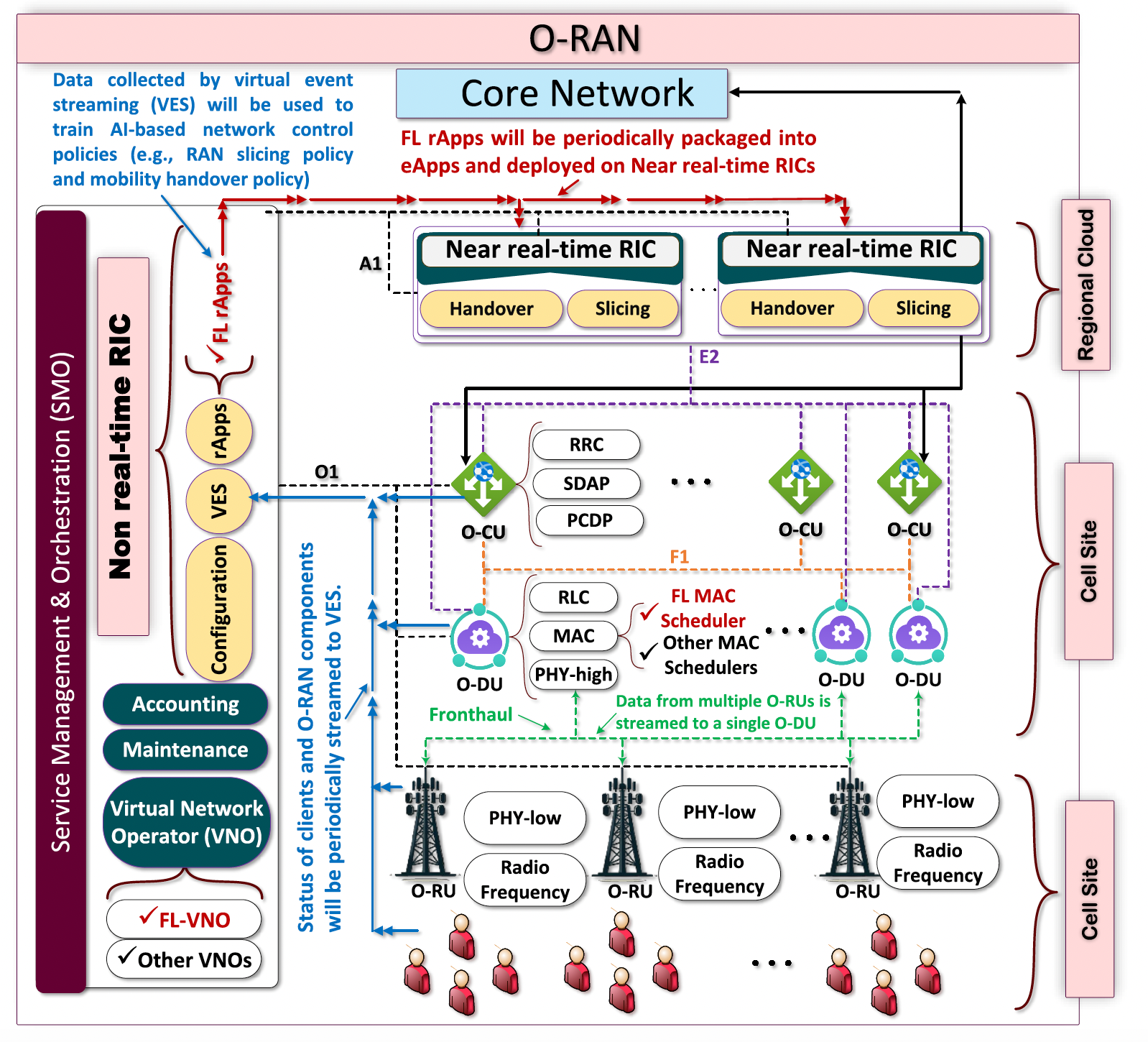}
        \vspace{-2mm}
    \caption{O-RAN architecture (all abbreviations follow standard 3GPP terminologies \cite{9846950}. O-RAN components interact via standard open interfaces (E2, F1, open fronthaul, A1, and O1), facilitating interoperability between network elements from different manufacturers.}
    \label{fig:ORANEvolution}
    \vspace{-4mm}
\end{figure*}

These considerations necessitate a revamp of the existing IoT network control functionalities to facilitate the coexistence of multiple FL-services within a dynamic network -- an approach absent in the literature. This naturally calls for a migration from
traditional RANs (e.g., distributed RANs in 4G), operating with black-box network control functionalities applied to all network services~\cite{9076124}.
\textit{To overcome this limitation and develop dedicated network control functionalities for FL-services, we leverage virtualization, intelligence, and programmability features of an emerging RAN technology used in 5G-and-beyond networks, called Open RAN.} 

O-RAN (Fig.~\ref{fig:ORANEvolution}) transitions from rigid cellular to multi-vendor and data-driven networks by integrating the concepts of intelligence, virtualization (RAN slicing), and programmable \textit{white-box} hardware (as opposed to the \textit{black-box} hardware)~\cite{9846950}. O-RAN supports multiple services with diverse QoS requirements via RAN slicing, which partitions RAN resources into isolated RAN slices, each leased to a virtual network operator (VNO)~\cite{9076124}. Unlike traditional RANs, O-RAN fosters dedicated functionalities for each type of service, a feature referred to as \textit{programmability}. For example, in contrast to the black-box MAC scheduler used in 4G that serves all services uniformly, O-RAN enables the implementation of a dedicated MAC scheduler for each service.

In this work, we exploit O-RAN's unique features, i.e., RAN slicing and programmability, and develop a novel intelligent FL architecture, called \textit{elastic FL ({\tt EFL})}. In {\tt EFL}, we introduce dedicated network control functionalities to orchestrate multiple FL-services over a shared dynamic wireless network.



\vspace{-3mm}
\subsection{Contributions}
Our major contributions are as follows:
\begin{itemize}[leftmargin=4mm]
    \item We identify unique challenges arising from the coexistence of multiple FL-services in dynamic environments. To address them, while considering the service-level objectives of FL-services,
    we propose \textit{elastic FL ({\tt EFL})} over O-RAN. 
    \item In {\tt EFL}, we use \textit{intelligence and programmability} features of O-RAN to implement dedicated dynamic network control functionalities for FL-services (relaxation of \textbf{(A1)}) at three time scales: non-real-time (non-RT), near-RT, and RT.
    \item At the non-RT level, we develop functionalities in {\tt EFL} that do not require immediate responsiveness, including client and FL-service registration, and training AI-based application, which we call estimator applications (eApps). eApps are utilized to predict FL-related factors, e.g., perspective-driven client distribution across O-RUs.
    \item At the near-RT level, we develop \textit{virtual RAN slices} for FL-services in {\tt EFL}, facilitating the seamless execution of multiple FL-services (relaxation of \textbf{(A2)}).
     {\tt EFL} also introduces a dedicated \textit{slice management unit} and a \textit{mobility management unit} to handle resource over-/under-provisioning and perspective-driven load balancing.
    \item At the RT level, we introduce dedicated FL MAC scheduler in {\tt EFL} to dynamically allocate resources to clients considering service-level objectives of FL-services.
\end{itemize}

\vspace{-3mm}
\section{{\tt EFL}: Elastic FL}\label{LsFedLOverORan}
\noindent In the following, we systematically introduce the {\tt EFL} framework, which will be implemented over O-RAN. 

\subsection{Preliminaries on O-RAN}
5G-and-beyond IoT networks are expected to support diverse services, including  ultra-reliable low latency communications and massive machine-type communications~\cite{9076124}. They also are envisioned to provide services for various \textit{verticals} -- groups of companies requiring similar services (e.g., industrial factories) -- operated by distinct \textit{virtual network operators}. This versatile service provisioning is enabled by the use of \textit{intelligent and programmable} O-RANs (Fig.~\ref{fig:ORANEvolution}). In O-RAN, 3GPP stack functionalities are disaggregated into (i) radio unit (O-RU), (ii) distributed unit (O-DU), and (iii) centralized unit (O-CU)~\cite{9846950}. O-RAN also introduces RAN intelligent controllers (RICs), i.e., non-real-time (non-RT) RIC and near-RT RICs, which utilize AI/ML algorithms to manage RAN operations at different time-scales~\cite{9846950,9796744}.

      \vspace{-3mm}
\subsection{Integration of {\tt EFL} in O-RAN}


\begin{figure*}
    \centering
    \includegraphics[width=0.91\linewidth,trim=1 1 1 1,clip]{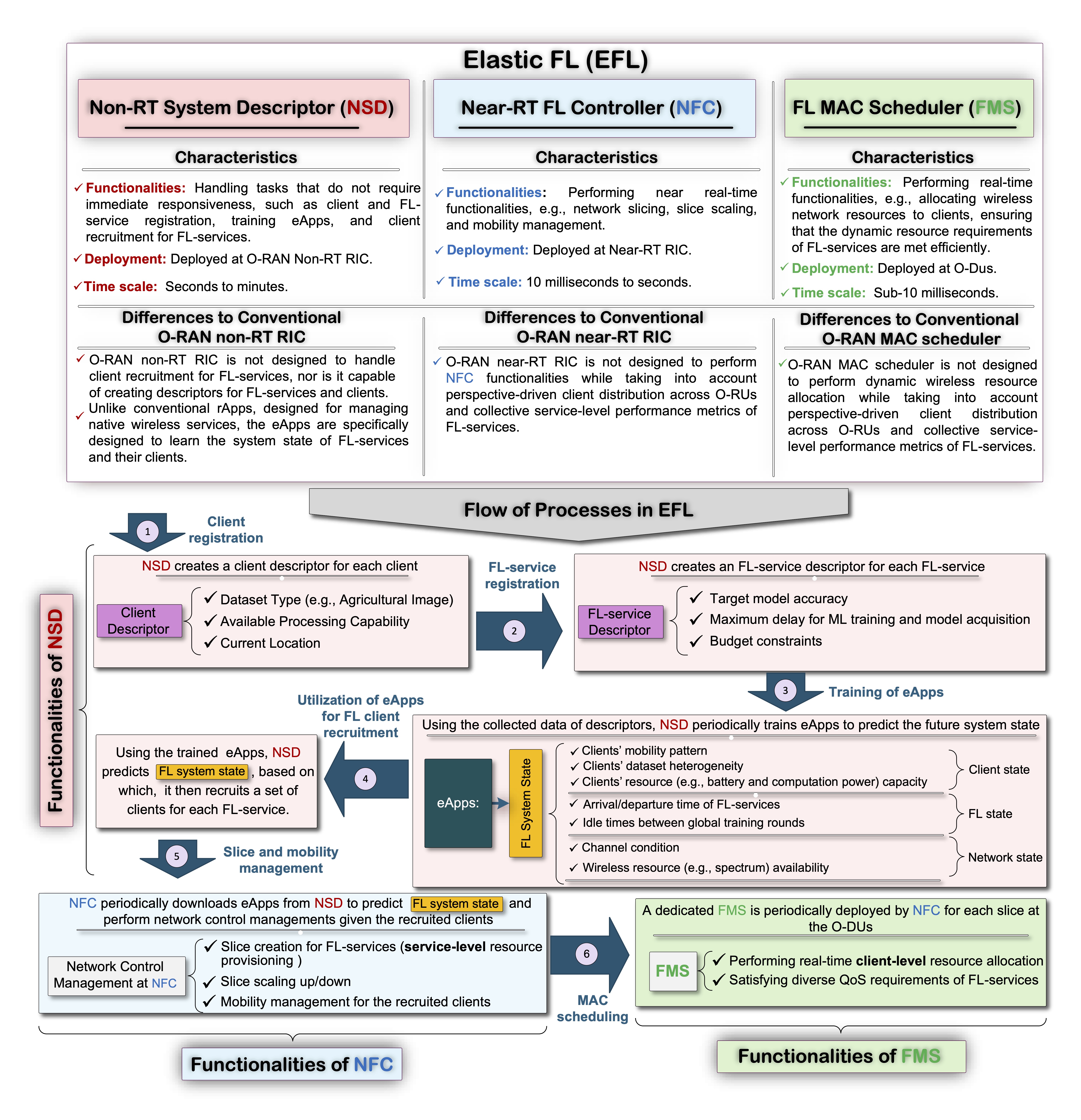}
    \vspace{-3mm}
    \caption{{ Top Panel: The three main components of EFL (NSD, NFC, and FMS) are illustrated along with their key characteristics and distinguishing features.
Bottom Panel: The process flow of EFL execution is depicted, showing the integration of NSD (red boxes), NFC (blue box), and FMS (green box). The sequence of operations is indicated by directional arrows, each annotated with a descriptive label written in blue text.}}
    \label{fig:EFL}
    \vspace{-0mm}
\end{figure*}

To integrate {\tt EFL} in O-RAN, we first focus on intelligent network control in O-RAN. O-RAN supports intelligent control across three levels of agility: (i) \textit{Non-RT RIC} operates on a timescale of seconds to minutes, establishing network management policies and hosting applications, known as rApps, which handle long-term functionalities like training AI models for network management. These trained AI models will be packaged into applications, referred to by xApps, and deployed on near-RT RIC. (ii)  \textit{Near-RT RIC}, operating on timescales of 10 milliseconds to  seconds, leverages xApps to forecast traffic flow and network dynamics. Using these forecasts, near-RT RIC manages load balancing, network slicing, and mobility management.
(iii) Finally, the \textit{MAC Scheduler}, located at the O-DUs, operates in sub-10-millisecond intervals, focusing on real-time resource allocation to the clients.

{\tt EFL}, depicted in Fig.~\ref{fig:EFL}, exploits O-RAN programmability and 
envisions dedicated functionalities at each of the aforementioned control layers to manage the concurrent execution of multiple FL-services (relaxation of \textbf{(A2)}) in a dynamic wireless environment (relaxation of \textbf{(A1)}).
We refer to these functionalities as: \textit{(i) Non-RT System Descriptor}, located at the non-RT RIC; \textit{(ii) Near-RT FL Controller}, deployed at the near-RT RIC; (iii) \textit{FL MAC Scheduler}, deployed at the O-DUs. We next design these functionalities to consider the collective service-level performance requirements, such as FL model accuracy across all FL-services. 

\vspace{-2mm}
\subsection{Non-RT System Descriptor} \label{ETS}

The coexistence of multiple FL-services introduces a new dimension of heterogeneity, which we call \textit{heterogeneity of interest}. This heterogeneity reflects the \textit{diverse interests} of FL-services in recruiting clients with different (i) datasets, (ii) data quality, (iii) channel conditions, and (iv) processing capabilities. These preferences are shaped by quality-of-service (QoS) requirements of each FL-service (e.g.,  target model accuracy and latency sensitivity) and its budget to recruit clients and acquire wireless resources.

To monitor the status of clients and FL-services, we equip {\tt EFL} with \textit{non-RT system descriptor}, deployed at the O-RAN non-RT RIC. This component handles long-term operations (ranging from seconds to minutes) such as client and FL-service registration, along with training AI-based applications for network control. Specifically, this unit stores and regularly updates two descriptors: (i) \textit{client descriptor}  and  (ii) \textit{FL-service descriptor}.
The client descriptor contains information about each client such as dataset type/size. The FL-service descriptor consists of key QoS requirements such as maximum delay for ML training and model acquisition, target model accuracy, and budget for client recruitment and wireless resource acquisition.  Using the information these descriptors provide, the non-RT system descriptor trains AI-based control applications, which we call estimator applications (eApps).


eApps can be construed as a special case of rApps, where due to their unique features, we have opted for a new name. The key difference between eApps and conventional rApps is their focus on \textit{service-level requirements unique to the FL}.
For example, client congestion in FL extends beyond simple network traffic volume; it also encompasses unique FL characteristics, including recruitment costs and dataset types (e.g., congestion of clients suitable for an object detection FL-service vs a product recommendation FL-service may differ across O-RUs). Subsequently, using eApps, {\tt EFL}  becomes capable of
satisfying the service-level requirements of FL-services -- a nuanced capability absent from existing FL methodologies. In particular, in {\tt EFL}, eApps are utilized to predict/estimate various system factors, including perspective-driven client distribution at O-RUs and fluctuations in the descriptors of clients and FL-services. 
These estimates enable {\tt EFL} to dynamically shift the resources between FL-services based on their real-time demands for clients and wireless resources.


\vspace{-2mm}
\subsection{Near-RT FL Controller}
To manage over-/under-provisioning of wireless resources and provide an environment for seamless execution of multiple FL-services, we equip {\tt EFL} with \textit{near-RT FL controllers}. A near-RT FL controller, operating on a timescale from 10 milliseconds to seconds, periodically downloads eApps to predict the future \textit{FL system state}, e.g., client congestion at O-RUs. Using these predictions, near-RT FL controller then performs functionalities that require faster response times than those handled by the non-RT system descriptor. These functionalities include \textit{FL slice creation}, \textit{elastic slice scaling up/down}, and \textit{client mobility management}, explained below.

\subsubsection{FL slice creation} 
{\tt EFL} enables the coexistence of multiple FL-services by creating a virtual RAN slice for each FL-service. 
In particular, using eApps, the near-RT FL controller first predicts the FL system states (e.g., client congestion at O-RUs). These predictions are derived from both current and historical data provided by the non-RT system descriptor. Based on these predictions, the near-RT FL controller then sends a \textit{slice creation request} for each FL-service to the near-RT RIC responsible for RAN slicing. This request specifies (i) the required resources, such as bandwidth/spectrum, and (ii) the QoS requirements for each FL-service, including delay constraints and priority for accessing O-RAN resources. The slice is then provisioned with the wireless resources (e.g., spectrum) and sent to O-DUs where FL MAC schedulers perform real-time resource allocation for clients of FL-services. 

For example, consider a time-sensitive FL-service (e.g., traffic prediction) with strict delay constraints for receiving local models from its clients. 
Additionally, presume that the datasets across the clients in this FL-service are highly heterogeneous.
This dataset heterogeneity leads to a higher demand for timely reception of the local models of large number of clients at the model aggregations of FL as missing clients' local models may prevent the global model from being exposed to certain client data, severely affecting its prediction performance. 
In this scenario, the near-RT FL controller in {\tt EFL} creates a dedicated RAN slice, with high-priority access to O-RAN resources, for this FL-service to ensure successful transmission of the local models of its clients.


\subsubsection{Elastic slice scaling down/up} 
To address resource over/under-provisioning, the near-RT FL controller performs elastic slice scaling down/up. This operation dynamically adjusts the wireless resources (e.g., spectrum) provisioned to slices based on the client congestion and resource demands of FL-services estimated by eApps. For example, if a high-priority, time-sensitive FL-service (e.g., traffic prediction) arrives with significant resource demands predicted by eApps, the near-RT FL controller may scale down an existing slice (i.e., reallocating its resources) to ensure that this FL-service receives enough resources to meet its QoS requirements.
\subsubsection{Client mobility management}
The near-RT FL controller periodically estimates the service-level resource requirements of each FL-service and assesses wireless channel quality -- such as data rates -- between clients and O-RUs using eApps. Based on these estimates, the near-RT FL controller alters the client to O-RU connections (i.e., conducts handovers). 
For example, assume that some clients of a time-sensitive FL-service (e.g., traffic prediction) move into the coverage area of an O-RU with better channel conditions than their current O-RUs. To maintain low-latency model transmission for these clients, the near-RT FL controller may prioritize the time-sensitive FL-service by shifting the connections of these clients to the target O-RU with stronger channel conditions. At the target O-RU, this operation may require reallocating resources from  the clients of other FL-services with lower time-sensitivity (e.g., decreasing their allocated bandwidth) to accommodate the clients of the higher-priority service.

\begin{figure*}
    \centering
    \includegraphics[width=0.6\linewidth]{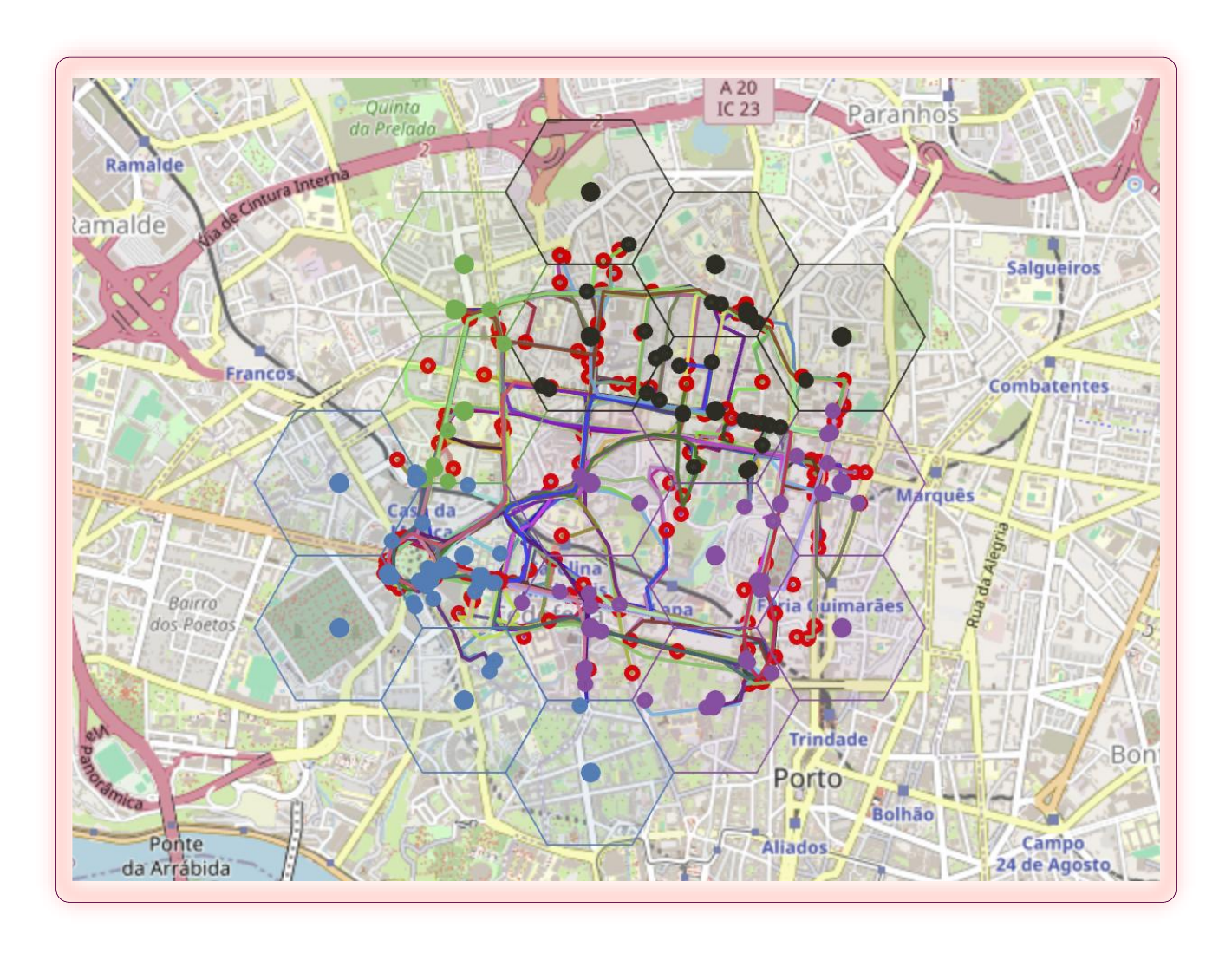}
    \vspace{-4mm}
    \caption{Real-world distribution of 150 FL clients over the Porto City Taxi Dataset: \href{https://www.kaggle.com/datasets/crailtap/taxi-trajectory/data}{https://www.kaggle.com/datasets/crailtap/taxi-trajectory/data}. Clients follow their actual mobility trajectories across the city and are served by $18$ O-RUs forming a realistic cellular network. These O-RUs are grouped under $4$ O-DUs, such that each O-DU serves on average $4$--$5$ O-RUs. The colored dots in the figure represent the locations of individual clients, where different colors indicate clients managed by the same O-DU. The coverage areas of the O-RUs, each spanning 500 meters, are shown as colored hexagonal cells, where different colors indicate O-RUs managed by the same O-DU. In addition, the trajectories of the clients are shown by lines across the city. This setup creates a realistic dynamic wireless environment for evaluating slice management, mobility handling, and MAC scheduling within the proposed EFL framework.}
    \label{fig:map_medium}
\end{figure*}

\begin{figure*}
\centering
          \includegraphics[width=0.75\linewidth,trim=10 1 10 1,clip]{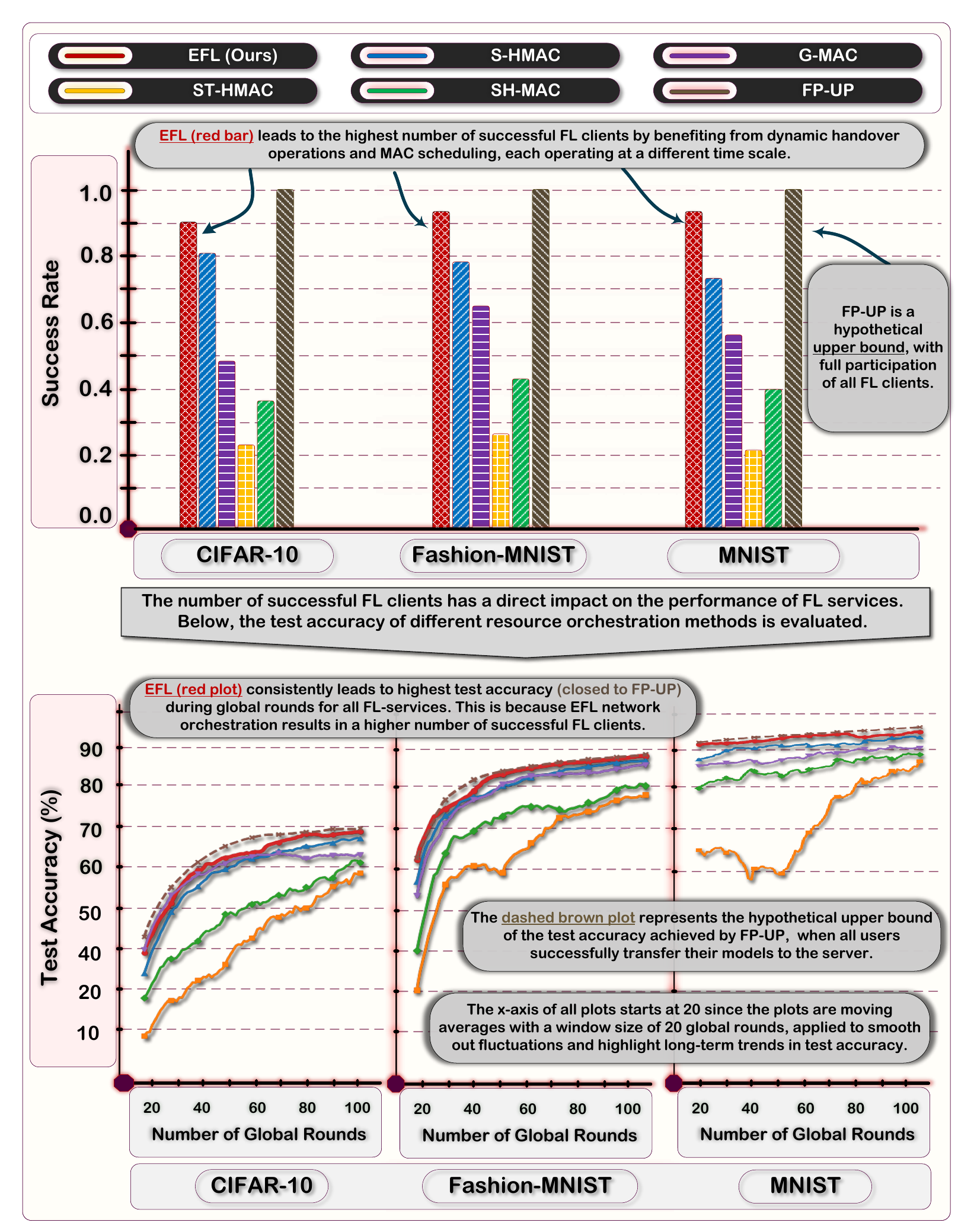}
           \vspace{-3.0mm}
         \caption{{The number of successful clients and its impact on the model prediction of FL-services under baseline methods and our proposed {\tt EFL} method. Our proposed method outperforms the baseline methods in terms of the number of successful clients and the model prediction performance of FL-services. {\tt EFL} leads to the success rate of above 0.9 for all the datasets, which is better than the best baseline (i.e., S-HMAC), which can reach the top success rate of 0.8. 
         Additional insights can be found in the annotated text within the plots.}}
         \label{fig:signal_fluctuation}
         \vspace{-0.5mm}
\end{figure*}

\vspace{-3mm}
\subsection{FL MAC Scheduler}

{\tt EFL} introduces a new dimension to MAC schedulers, enabling client-level, real-time resource allocation for FL clients. This addresses resource over/under-provisioning at a shorter time-scale compared to near-RT FL controller, tailored to service-level performance metrics of FL-services, such as delay constraints for receiving local models from clients. Specifically, for FL-services, we equip {\tt EFL} with a \textit{dedicated FL MAC scheduler} located at O-DU to allocate wireless resources to FL clients. 

Resource allocation by the FL MAC scheduler is different from resource provisioning by the near-RT FL controller, as the FL MAC scheduler manages \textit{client-level} resource allocation (e.g., spectrum allocation to each client), responding to the real-time demands of individual clients. Conversely, the near-RT FL controller focuses on \textit{service-level} resource provisioning (e.g., spectrum allocation to each slice), considering the broader resource needs of each FL-service, which typically change  slower than rapid variations in client-level resource demands (e.g., their needed bandwidth which exhibits a high temporal variation caused by the channel fluctuations).

      \vspace{-3mm}
\subsection{System Operation}
Fig.~\ref{fig:EFL} illustrates the integration of our proposed {\tt EFL} into O-RAN. At the top, the characteristics of different {\tt EFL} components are highlighted, along with their differences from conventional O-RAN functionalities. At the bottom, the figure depicts how  {\tt EFL} components interact.

{
\subsection{Challenges of Integration into the Existing O-RAN Implementations}
Integrating {\tt EFL} into existing commercial O-RAN implementations such as the OpenAirInterface (OAI) and Software-Defined RAN (SD-RAN) \cite{11124199} poses several challenges due to their current design limitations. First, both OAI and SD-RAN provide limited built-in support for FL-specific service-level objectives, such as model convergence time, client heterogeneity, and multi-service orchestration. Extending them to support our {\tt EFL} framework would require custom modifications to their RIC platforms to support estimator applications (eApps) and FL-specific slicing logic. Second, achieving real-time responsiveness across non-RT, near-RT, and RT layers necessitates tight synchronization and lightweight cross-layer APIs, which may not be natively supported in current RIC/xApp/rApp pipelines. For example, deploying FL MAC schedulers at the O-DU level would require modification of the real-time scheduler in OAI or SD-RAN’s platform to accommodate FL client-specific metrics such as local model delivery deadlines. Third, interoperability between vendor-specific O-RUs and generalized FL slice descriptors remains a challenge due to heterogeneous support for open fronthaul. Despite these challenges, the modular nature of OAI and SD-RAN makes them promising platforms for incremental integration of {\tt EFL}. 
}



      \vspace{-1mm}

\section{Case Study and Prototyping}\label{motivatingNewArchitecture} 
\noindent {For prototyping, we consider a dense urban O-RAN scenario with vehicular clients in Porto city. The positions and mobility trajectories of the clients are drawn from the real-world Porto city taxi dataset: \href{https://www.kaggle.com/datasets/crailtap/taxi-trajectory/data}{https://www.kaggle.com/datasets/crailtap/taxi-trajectory/data}. We model an O-RAN deployment with 4 O-DUs, 18 O-RUs, and 150 FL clients (see Fig.~\ref{fig:map_medium}). The O-RUs provide coverage over a hexagonal grid with a radius of 500 m, each assumed to have 5 MHz of available bandwidth.} We also assume that FL clients have a maximum transmit power of 300 mW. The O-RAN environment, including signal forms, channels, and data rates, is modeled according to~\cite{9771187}. { It is worth noting that the optimization problems in our framework are solved independently for each O-DU and O-RU in parallel, enabling scalability to accommodate large numbers of clients.}

{ We consider three FL-services, each recruiting $50$ clients, which utilize convolutional neural networks (CNNs) to train on CIFAR10 (FL-service 1), Fashion-MNIST (FL-service 2), and MNIST (FL-service 3) datasets~\cite{9492755}. In FL-service 1, data points are distributed based on Dirichlet distribution $\mathsf{Dir}(\alpha)$ \cite{10130784} with parameter $\alpha{=}0.1$, making clients' datasets highly heterogeneous. For FL-service 2 and FL-service 3, we use $\alpha{=}10$.}
Each FL service runs for 100 global training rounds, where global aggregations are performed using the FedAvg method~\cite{hosseinalipour2023parallel}. We assume that FL-service 1 has a strict delay constraint of $15$ seconds for receiving local models from its clients, while FL services 2 and 3 have a more relaxed delay constraint of $20$ seconds. Clients that fail to fully transfer their models to their O-RUs within the designated time window are considered \textit{unsuccessful}.

 {$\star$ The details of our implementations, including all source codes and extended discussions on the mathematical formulations, are provided in our GitHub repository:
\href{https://github.com/payamsiabd/Elastic_FL_Over_O_RAN.git}{https://github.com/payamsiabd/Elastic\_FL\_Over\_O\_RAN.git}
}




\begin{figure*}
    \centering
    \includegraphics[width=0.75\linewidth,trim=10 1 10 1,clip]{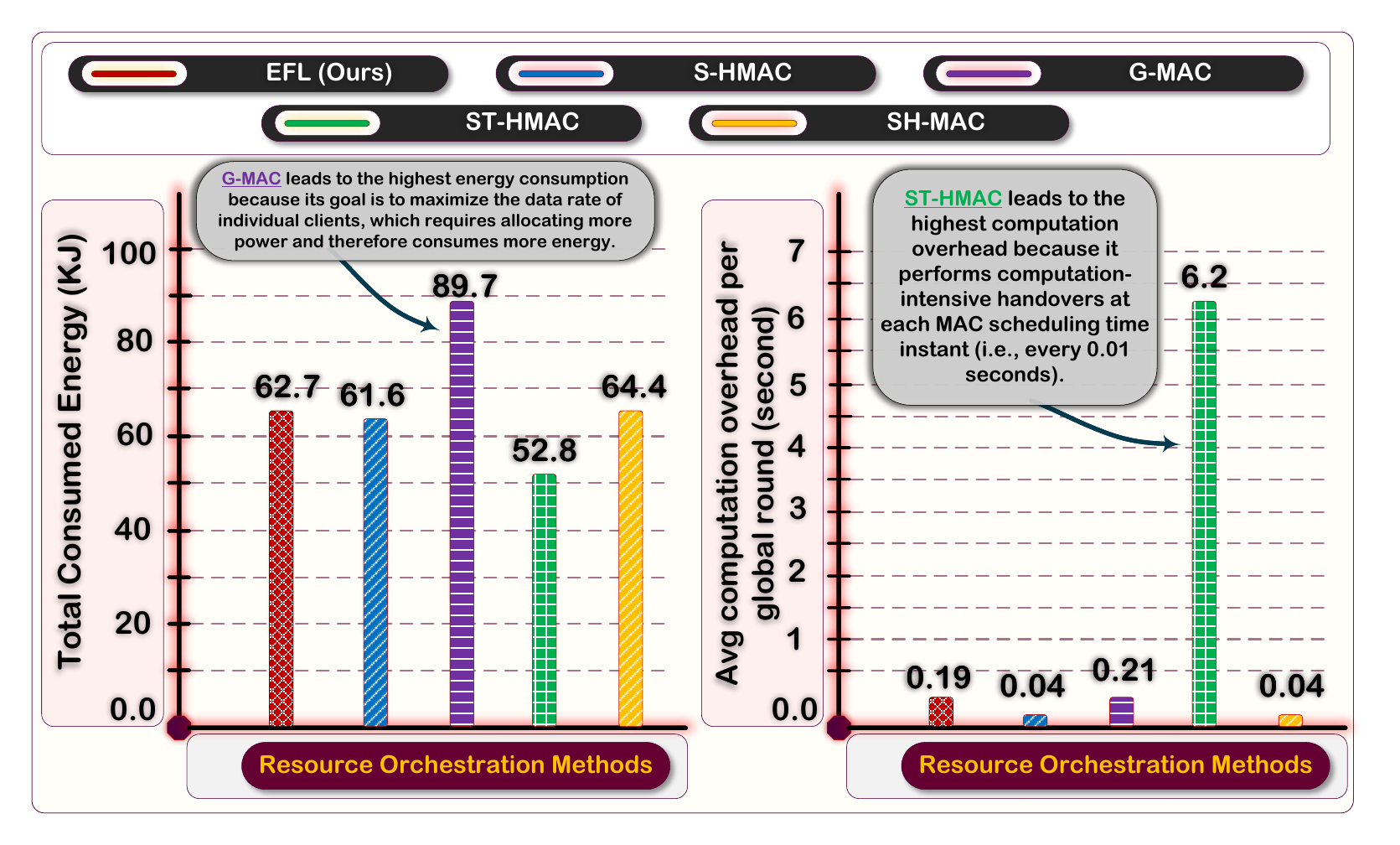}
    \vspace{-4mm}
    \caption{{Energy consumption and computation overhead of different methods. Our {\tt EFL} method maintains a low energy consumption and computation overhead compared to the baselines with the added benefit of model performance gains as reflected in Fig.~\ref{fig:signal_fluctuation}. The  numbers that correspond to each bar are represented above it for clarity. Additional insights can be found in the annotated text within the plots.}}
    \label{fig:RF_computation_energy}
    \vspace{-0.5mm}
\end{figure*}

\begin{figure*}
    \centering
    \includegraphics[width=0.75\linewidth,trim=10 1 10 1,clip]{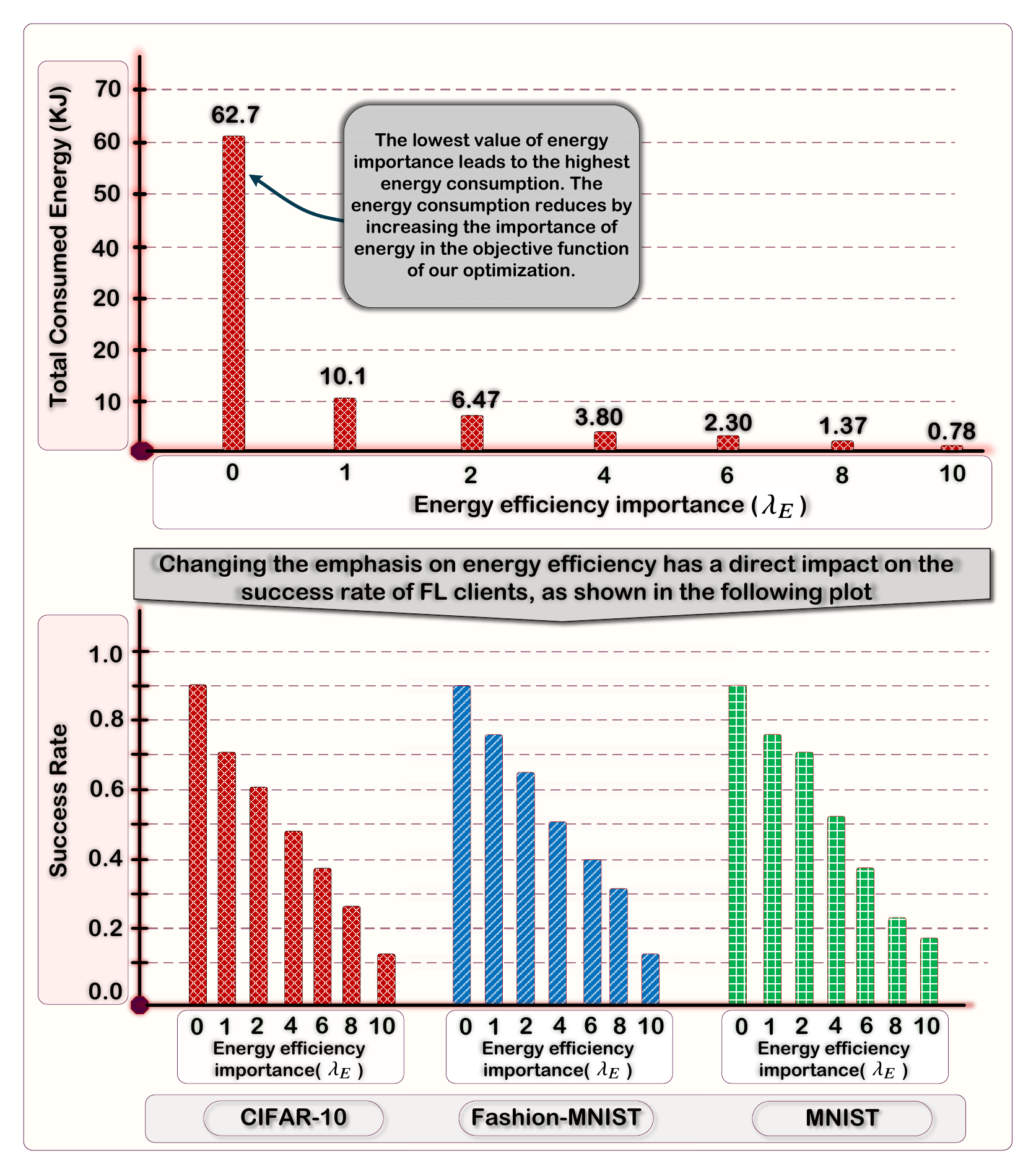}
    \vspace{-4mm}
    \caption{{Energy consumption and the number of successful clients of {\tt EFL} upon varying the importance of energy efficiency $\lambda_E$ in the objective function of our optimization. Increasing $\lambda_E$ leads to energy savings (the  numbers that correspond to each bar are represented above it for clarity) but it results in a drop in the number of successful clients, which in turn causes reductions in the model performance as reflected in Fig.~\ref{fig:signal_fluctuation}. This reveals the tradeoff between energy savings and model performance. Additional insights can be found in the annotated text within the plots.}}
    \label{fig:RF_lambda}
    \vspace{-0.5mm}
\end{figure*}

\vspace{-2mm}
\subsection{Prototyping of {\tt EFL} Components} We implement the core components of {\tt EFL} following the outlined methodology in Fig.~\ref{fig:EFL} as follows.
\subsubsection{Non-RT system descriptor}
To prototype this component, we first focus on arrows numbered 1 and 2 at the bottom of Fig.~\ref{fig:EFL} and implement features for storing client descriptor (containing clients' locations and dataset types) and FL-service descriptors (containing delay constraints for receiving local models). Considering the arrow numbered 3 in Fig.~\ref{fig:EFL},  we develop an eApp for mobility pattern prediction. 
To train the eApp, we run a simulation where {\tt EFL} starts with an untrained mobility prediction model, which uses past clients' positions to predict their future locations for the next 0.01 seconds (the model architecture is provided in the above GitHub). The model optimizes the mean squared error (MSE) loss function to minimize prediction errors in mobility patterns.  
{ We note that our eApps train rapidly (within a few learning iterations; see our GitHub for details) and provide accurate location predictions for the next 0.01 seconds. As a result, the impact of estimation errors is negligible, making our optimizations robust to such inaccuracies.}

Following the arrow numbered 4 in Fig.~\ref{fig:EFL}, the client recruitment for each FL-service is conducted based on the dataset type of each client (50 clients with CIFAR10 dataset are assigned to FL-service 1, 50 clients with Fashion-MNIST dataset are assigned to FL-service 2, and 50 clients with MNIST dataset are assigned to FL-service 3). 
Referring to Fig.~\ref{fig:EFL}, following the arrow number 5, the trained eApp is then used as a component in the near-RT FL controller to predict the system state, based on which network control managements are performed as follows.

\subsubsection{Near-RT FL controller}
Referring to Fig.~\ref{fig:EFL}, following the arrow number 5, we implement functionalities of near-RT FL controller for slice and mobility management jointly as an optimization problem. We define the operational time window for the near-RT FL controller as \textit{400 milliseconds}: every 400 milliseconds, the near-RT FL controller first leverages the eApp to predict client locations for the next 10 milliseconds and then estimates the data rate between clients and O-RUs. 
{ Near-RT FL controller then solves an optimization problem to determine the optimal transmit power of clients and the bandwidth allocated to each client to maximize $\sum_{s=1}^3  \delta_s - \lambda_E \sum_{u} P_u$, which represents the difference between two terms: (i) an FL-service–level metric, formulated as the sum of successful clients $\delta_s$ of each FL-service $s$, and (ii) the total energy consumption of all clients $\sum_{u} P_u$. Here, $\lambda_E$ specifies the importance assigned to energy efficiency, and $P_u$ denotes the transmit power of client $u$. 
Detailed formulation of our optimization problem are provided in the above GitHub.}

Near-RT FL controller then creates a slice creation request for each FL-service that specifies (i) the maximum transmit power allowed for each client within the slice and (ii) the set of FL clients connected to each O-RU, defined as those whose allocated power exceeds a small threshold of $0.01$ W. It then sends the request to O-RAN near-RT RIC to create a RAN slice at each O-RU for each FL-service. 


\subsubsection{FL MAC scheduler} 
Referring to Fig.~\ref{fig:EFL}, and following arrow number 6, we design FL MAC schedulers that operate in real-time (every 10 milliseconds) using information from the near-RT FL controller. The MAC schedulers leverage eApps to predict client locations over the next 10 milliseconds and, based on these predictions, estimate the channel gains between clients and their connected O-RUs. They then solve lightweight optimization problems to allocate bandwidth and power to clients. These optimization problems follow the same structure as the one solved by the near-RT controller but differ in two key ways: (i) they are solved \emph{per O-RU} and only over the clients currently connected to that O-RU (as determined by the near-RT controller), and (ii) the \emph{client power on each link} is upper bounded by the \emph{slice-level power cap} computed by the near-RT controller.




{  We compare {\tt EFL} against five methods:

\begin{enumerate}
    \item \textbf{Single time-scale dynamic handover and MAC scheduling (ST-HMAC):} both handover and MAC scheduling are performed at each MAC scheduling interval (i.e., every 0.01 s), inspired by \cite{kang2023joint}.
    \item  \textbf{Snapshot handover and dynamic MAC scheduling (S-HMAC):} handover is performed only once at the start of each global round, while MAC scheduling is updated dynamically every 0.01 s, inspired by \cite{feng2022mobility}.
    \item  \textbf{Snapshot handover and MAC scheduling (SH-MAC):} inspired by conventional FL orchestrators~\cite{9445589,hosseinalipour2023parallel,9799768}, this method performs both handover and MAC scheduling once at the beginning of each global round.
    \item  \textbf{General MAC scheduling (G-MAC):} resembling current MAC schedulers used in general internet services such as video streaming~\cite{9076124,8737604,9001216}, this method does not incorporate FL-specific QoS considerations such as the number of successful clients or prioritization among FL services.
    \item  \textbf{Full participation upper bound (FP-UP):} an idealized benchmark where all FL clients in all services are always able to successfully transmit their local models to the server, serving as a hypothetical upper bound on test accuracy (due to violating the resource constraints).
\end{enumerate}}

\subsection{Results and Discussions}

The top plot in Fig.~\ref{fig:signal_fluctuation} shows that ST-HMAC yields the worst performance as it performs handover operations and MAC scheduling every 0.01 seconds, which introduces substantial overhead. Specifically, since FL clients must upload their local models to the server within a strict deadline, the time wasted by these frequent control operations reduces the available transmission window, thereby increasing the number of unsuccessful clients. This is further reflected by its worst performance across the methods in the bottom plot of Fig.~\ref{fig:signal_fluctuation}. Also, it can be seen that S-HMAC performs better than the other baselines but worse than our {\tt EFL}. This is because, although S-HMAC carries out dynamic MAC scheduling, it still performs handover operations only at the start of each global round (the initially optimal connections between clients and O-RUs may become suboptimal due to client mobility). Moreover, G-MAC results in more failure of clients compared to {\tt EFL} as it focuses on maximizing each client's data rate (i.e., a per-client performance metric), rather than maximizing the number of successful clients for FL-services (i.e., service-level performance metric). This highlights how MAC-layer resource allocation and mobility management designed for native wireless services fall short in addressing the unique demands of FL-services.
As illustrated in Fig.~\ref{fig:signal_fluctuation}, our {\tt EFL} outperforms all baselines by enabling more clients to successfully transmit their local models to O-RUs. This is further reflected by its resulting model accuracies that are comparable to the upper bound achieved by FP-UP.


The left plot of Fig.~\ref{fig:RF_computation_energy} depicts the energy consumption of all methods. G-MAC consumes the most energy as its objective prioritizes maximizing user data rates, which in turn drives higher power consumption. The energy consumption of {\tt EFL} is comparable to ST-HMAC, S-HMAC, and SH-MAC, demonstrating that our framework does not impose additional energy burdens. Importantly, while maintaining similar energy costs, {\tt EFL} yields the highest model accuracy among all algorithms, as discussed above, by enabling the largest number of successful clients to transmit their models within the deadline. 
Also, the right plot in Fig.~\ref{fig:RF_computation_energy} reports the average computation time overhead per global round incurred by handover operations and MAC scheduling. 
{\tt EFL} achieves a computation overhead that is close to the best baselines; however, it is worth noting that 
S-HMAC and SH-HMAC yield lower computation overheads compared to {\tt EFL} as they perform handover operation only once at the beginning of each global round. Nevertheless, these additional overheads of {\tt EFL} remain tractable and are justified by the gains in the previously discussed client success rate and model accuracy. This highlights the advantage of our design in balancing computational overhead, energy efficiency, and FL accuracy.

In Fig.~\ref{fig:RF_lambda}, we conduct analyses on the value of $\lambda_E$, which tunes the importance of energy term in our optimization, to show how varying this coefficient affects both the success rate of clients and the total energy consumption of {\tt EFL}. As shown, by increasing the value of $\lambda_E$, our method prioritizes energy savings (i.e., it reduces the transmit power of clients) by sacrificing the number successful clients, which will lead to a drop in model accuracy as we studied in Fig.~\ref{fig:signal_fluctuation}.
These results highlight the tradeoff between energy conservation and client participation, offering insights into how EFL can be tailored to energy-constrained IoT devices depending on the prioritized objective (model performance vs. energy efficiency).

\section{Conclusion and Future Work}\label{conclusion}
\noindent
We proposed {\tt EFL}, an innovative FL architecture over O-RAN, orchestrating the concurrent execution of multiple FL-services in a dynamic wireless network. In {\tt EFL}, we introduced three layers of dedicated network control functionalities for FL-services: the non-RT system descriptor, near-RT FL controller, and FL MAC scheduler. 
Additionally, we implemented a prototype of {\tt EFL} and demonstrated the notable performance gains it can obtain compared to the baselines methods.


The introduction of {\tt EFL} opens the door to various future innovations, including creating optimized slice elasticity units that balance client energy usage, O-RAN operational costs, and the efficiency of FL-service model training. Also, addressing adversarial behaviors — such as FL-services misrepresenting their resource requirements — is crucial, as these can negatively impact overall system performance. Further, considering direct device-to-device communications between the clients opens up a new research avenue on designing optimal RICs and MAC schedulers considering ML model relaying between the clients. 







\bibliographystyle{IEEEtran}
\bibliography{OFedRefs}

\begin{thebibliography}{10}
\providecommand{\url}[1]{#1}
\csname url@samestyle\endcsname
\providecommand{\newblock}{\relax}
\providecommand{\bibinfo}[2]{#2}
\providecommand{\BIBentrySTDinterwordspacing}{\spaceskip=0pt\relax}
\providecommand{\BIBentryALTinterwordstretchfactor}{4}
\providecommand{\BIBentryALTinterwordspacing}{\spaceskip=\fontdimen2\font plus
\BIBentryALTinterwordstretchfactor\fontdimen3\font minus \fontdimen4\font\relax}
\providecommand{\BIBforeignlanguage}[2]{{%
\expandafter\ifx\csname l@#1\endcsname\relax
\typeout{** WARNING: IEEEtran.bst: No hyphenation pattern has been}%
\typeout{** loaded for the language `#1'. Using the pattern for}%
\typeout{** the default language instead.}%
\else
\language=\csname l@#1\endcsname
\fi
#2}}
\providecommand{\BIBdecl}{\relax}
\BIBdecl

\bibitem{9060868}
W.~Y.~B. Lim, N.~C. Luong, D.~T. Hoang, Y.~Jiao, Y.-C. Liang, Q.~Yang, D.~Niyato, and C.~Miao, ``Federated learning in mobile edge networks: A comprehensive survey,'' \emph{IEEE Commun. Surveys \& Tuts.}, vol.~22, no.~3, pp. 2031--2063, 2020.

\bibitem{9799768}
Z.~Cheng, M.~Liwang, X.~Xia, M.~Min, X.~Wang, and X.~Du, ``Auction-promoted trading for multiple federated learning services in {UAV}-aided networks,'' \emph{IEEE Trans. Veh. Technol.}, vol.~71, no.~10, pp. 10\,960--10\,974, 2022.

\bibitem{9445589}
M.~N.~H. Nguyen, N.~H. Tran, Y.~K. Tun, Z.~Han, and C.~S. Hong, ``Toward multiple federated learning services resource sharing in mobile edge networks,'' \emph{IEEE Trans. Mobile Comput.}, vol.~22, no.~1, pp. 541--555, 2023.

\bibitem{9076124}
J.~Li, W.~Shi, P.~Yang, Q.~Ye, X.~S. Shen, X.~Li, and J.~Rao, ``A hierarchical soft {RAN} slicing framework for differentiated service provisioning,'' \emph{IEEE Wireless Commun.}, vol.~27, no.~6, pp. 90--97, 2020.

\bibitem{8737604}
S.~Mandelli, M.~Andrews, S.~Borst, and S.~Klein, ``Satisfying network slicing constraints via {5G} {MAC} scheduling,'' in \emph{IEEE INFOCOM}, 2019, pp. 2332--2340.

\bibitem{9846950}
S.~D'Oro, M.~Polese, L.~Bonati, H.~Cheng, and T.~Melodia, ``d{A}pps: Distributed applications for real-time inference and control in {O-RAN},'' \emph{IEEE Commun. Mag.}, vol.~60, no.~11, pp. 52--58, 2022.

\bibitem{9796744}
S.~D’Oro, L.~Bonati, M.~Polese, and T.~Melodia, ``Orchest{RAN}: Network automation through orchestrated intelligence in the open {RAN},'' in \emph{IEEE INFOCOM}, 2022, pp. 270--279.

\bibitem{11124199}
K.~Alam, M.~A. Habibi, M.~Tammen, D.~Krummacker, W.~Saad, M.~D. Renzo, T.~Melodia, X.~Costa-Pérez, M.~Debbah, A.~Dutta, and H.~D. Schotten, ``A comprehensive tutorial and survey of o-ran: Exploring slicing-aware architecture, deployment options, use cases, and challenges,'' \emph{IEEE Commun. Surv. Tutor.}, pp. 1--1, 2025.

\bibitem{9771187}
S.~Mondal and M.~Ruffini, ``Optical front/mid-haul with open access-edge server deployment framework for sliced {O-RAN},'' \emph{IEEE Trans. Netw. Service Manag.}, vol.~19, no.~3, pp. 3202--3219, 2022.

\bibitem{9492755}
J.~Mills, J.~Hu, and G.~Min, ``Multi-task federated learning for personalised deep neural networks in edge computing,'' \emph{IEEE Trans. Parallel Distrib. Syst.}, vol.~33, no.~3, pp. 630--641, 2022.

\bibitem{10130784}
H.~Li, Z.~Cai, J.~Wang, J.~Tang, W.~Ding, C.-T. Lin, and Y.~Shi, ``{FedTP}: Federated learning by transformer personalization,'' \emph{IEEE Trans. Neural Netw. Learn. Syst.}, vol.~35, no.~10, pp. 13\,426--13\,440, 2024.

\bibitem{hosseinalipour2023parallel}
S.~Hosseinalipour, S.~Wang, N.~Michelusi, V.~Aggarwal, C.~G. Brinton, D.~J. Love, and M.~Chiang, ``Parallel successive learning for dynamic distributed model training over heterogeneous wireless networks,'' \emph{IEEE/ACM Trans. Netw.}, 2023.

\bibitem{kang2023joint}
Y.~Kang, Y.~Zhu, D.~Wang, Z.~Han, and T.~Ba{\c{s}}ar, ``Joint server selection and handover design for satellite-based federated learning using mean-field evolutionary approach,'' \emph{IEEE Trans. Netw. Sci. Eng.}, vol.~11, no.~2, pp. 1655--1667, 2023.

\bibitem{feng2022mobility}
C.~Feng, H.~H. Yang, D.~Hu, Z.~Zhao, T.~Q. Quek, and G.~Min, ``Mobility-aware cluster federated learning in hierarchical wireless networks,'' \emph{IEEE Trans. Wireless Commun.}, vol.~21, no.~10, pp. 8441--8458, 2022.

\bibitem{9001216}
S.~Deng, Z.~Xiang, P.~Zhao, J.~Taheri, H.~Gao, J.~Yin, and A.~Y. Zomaya, ``Dynamical resource allocation in edge for trustable internet-of-things systems: A reinforcement learning method,'' \emph{IEEE Trans. Indust. Inform.}, vol.~16, no.~9, pp. 6103--6113, 2020.

\end{thebibliography}
\vspace{-13mm}
\begin{IEEEbiographynophoto}{Payam Abdisarabshali [S]} received the M.Sc. degree in Computer Engineering from Razi University, Iran, with top-rank recognition in 2018. He was a teaching assistant professor at Razi University from 2018 to 2022. He is currently a Ph.D. student at the Department of Electrical Engineering, University at Buffalo--SUNY, NY, USA. His research interests include distributed and federated machine learning, computer vision, the design of neural network architectures, and mathematical modeling.  
\end{IEEEbiographynophoto}
\vspace{-13.5mm}
\begin{IEEEbiographynophoto}{Nicholas Accurso [S]} received the M.Sc. degree from University at Buffalo (SUNY). He is currently a PhD student at  Department of Electrical Engineering of the University at Buffalo (SUNY). 
\end{IEEEbiographynophoto}
\vspace{-13.5mm}
\begin{IEEEbiographynophoto}{Filippo Malandra [M]} received the B.Eng. and M.Eng. degrees in telecommunications engineering from the Politecnico di Milano, Milan, Italy, in 2008 and 2011, respectively, and the Ph.D. degree in electrical engineering from Polytechnique Montreal, Montreal, QC, Canada, in 2016. He is an Assistant Professor with the Department of Electrical Engineering, State University of New York at Buffalo, Buffalo, NY, USA, where he is the Director of the Wireless Network for Smart Systems (WN4SS) Lab. His research interests include the performance analysis and design of mobile networks for smart cities, smart grid IoT applications, and the design of new solutions to provide broadband Internet access to underserved communities.

\end{IEEEbiographynophoto}
\vspace{-13mm}
\begin{IEEEbiographynophoto}{Weifeng Su [F]} received the B.S. and Ph.D. degrees in mathematics from Nankai University, Tianjin, China, in 1994 and 1999, respectively,
and the Ph.D. degree in electrical engineering from the University of Delaware, Newark, USA, in 2002. He is currently a Professor with the Department of Electrical Engineering, State University of New York (SUNY), Buffalo, USA. He has authored more than 100 peer-reviewed papers in IEEE journals and leading international conferences. He has coauthored the book Cooperative Communications and Networking (Cambridge University Press, 2009). His research interests span a broad range of areas from signal processing to wireless communications and networking, including space-time coding and modulation for MIMO wireless communications, MIMO-OFDM systems, and cooperative communications for wireless networks.

\end{IEEEbiographynophoto}
\vspace{-13mm}
\begin{IEEEbiographynophoto}{Seyyedali Hosseinalipour [SM]} received the B.S. degree in electrical engineering from Amirkabir University of Technology, Tehran, Iran, in 2015 with high honor and top-rank recognition. He then received the M.S. and Ph.D. degrees in electrical engineering from North Carolina State University, NC, USA, in 2017 and 2020, respectively; and was a postdoctoral researcher at Purdue University, IN, USA from 2020 to 2022. He was the recipient of the \textit{ECE Doctoral Scholar of the Year Award} (2020) and \textit{ECE Distinguished Dissertation Award} (2021) at NC State University; and \textit{Students’ Choice Teaching Excellence Award} (2023) at University at Buffalo--SUNY. Furthermore, he was the first author of a paper published in IEEE/ACM Transactions on Networking that received the \textit{2024 IEEE Communications Society William Bennett Prize}.
He has served as the TPC Co-Chair of workshops/symposiums related to machine learning and edge computing for IEEE INFOCOM, GLOBECOM, ICC, CVPR, ICDCS, SPAWC, WiOpt, and VTC. He has also served as the guest editor of IEEE Internet of Things Magazine for the special issue on \textit{Federated Learning for Industrial Internet of Things} (2023). Since Feb. 2025, he has been serving as an \textit{Associate Editor} for the \textit{IEEE Transactions on Signal and Information Processing over Networks}.
His research interests include the analysis of
modern wireless networks, synergies between machine learning methods and fog/edge

\end{IEEEbiographynophoto}
\end{document}